\documentclass[conference]{IEEEtran}

\newcommand{\nxbft}[0]{NxBFT}
\newcommand{\nxb}[0]{\textsc{NxB}}
\newcommand{\quorum}{\lfloor \frac{n}{2} \rfloor + 1}
\newcommand{\kops}{\,\mathrm{kOp}/\mathrm{s}}
\newcommand{\s}{\,\mathrm{s}}
\newcommand{\ms}{\,\mathrm{ms}}

\usepackage{amsmath,amssymb,amsfonts}
\usepackage{textcomp}
\usepackage{xcolor}
\usepackage{comment}

\usepackage{graphicx}
\usepackage{hyperref}
\usepackage{cleveref}
\usepackage{subfigure}
\usepackage{makecell}
\usepackage{multirow}
\usepackage{tikz}

\usepackage[ruled]{algorithm}
\usepackage[noend]{algpseudocode}
\algdef{SL}[REPLICASTATE]{ReplicaState}{0}[1]{\textbf{state} #1}
\algblockdefx[FN]{Fn}{EndFn}%
[3]{\textbf{function} \emph{#1}(#2) : #3}{}
\algblockdefx[PROC]{Proc}{EndProc}%
[2]{\textbf{function} \emph{#1}(#2)}{}
\algblockdefx[UPON]{Upon}{EndUpon}%
[2]{\textbf{upon} \texttt{#1} (#2)}{}
\algtext*{EndUpon}
\algtext*{EndFn}
\algtext*{EndProc}

\begin{document}

\title{Not eXactly Byzantine: Efficient and Resilient TEE-Based State Machine Replication}

\author{\IEEEauthorblockN{Marc Leinweber, Hannes Hartenstein}
\IEEEauthorblockA{KASTEL Security Research Labs \\
Karlsruhe Institute of Technology (KIT)\\
marc.leinweber@kit.edu, hannes.hartenstein@kit.edu}
}
\maketitle

\thispagestyle{plain}
\pagestyle{plain}

\begin{abstract}
We propose, implement, and evaluate \nxbft, a resilient and efficient State Machine Replication protocol using Trusted Execution Environments (TEEs). 
\nxbft{} focuses on a ``Not eXactly Byzantine'' (\nxb) operating model as a middle ground between crash and Byzantine fault tolerance. 
\nxbft's consensus layer is asynchronous, graph-based, leaderless, and optimized for the \nxb{} operating model, enabling load-balancing of requests between replicas and, in fault-free cases, two network round trips between decisions.
We identify fundamental issues with crash recovery due the use of TEEs in asynchrony that only can be circumvented by relying on synchrony for liveness.
We provide a throughput-latency trade-off analysis of \nxbft{}, Chained-Damysus (rotating leader), and MinBFT (static leader) for up to 40 replicas and network round trip latencies up to 150\,ms.
\nxbft{} achieves the highest throughput in all scenarios.
When small latencies are required, MinBFT and Damysus are at an advantage with Damysus benefiting from the \nxb{} model in terms of throughput for small deployments.
In contrast to leader-based approaches, \nxbft's performance is almost not impacted when actual crash faults occur.
\end{abstract}

\begin{IEEEkeywords}
Directed Acyclic Graph (DAG), SMR, Atomic Broadcast, Trusted execution, Asynchrony, Recovery
\end{IEEEkeywords}

\section{Introduction}

In this paper, we focus on federations in the permissioned blockchain model, where a set of operators (federation members) are running a State Machine Replication (SMR) protocol \cite{DBLP:journals/csur/Schneider90} to provide a common service. We assume that the federation consists, at most, of tens of operators and typically is spread over a country or continent. We also assume a main motivation of this operating model to be sovereignty reasons, that is, to have equal power and rights. Still, we assume that the federation members run a Byzantine atomic broadcast for resilience reasons. However, we deviate from classical Byzantine fault tolerant (BFT) SMR by assuming that a replica does not show Byzantine behavior in its interaction with its clients. 
We consider this assumption valid for use cases where the federation members are highly regulated and manipulation towards the client is directly observable and correctable, for example when logging check-ins and check-outs as part of a ticketing system for mobility-as-a-service federation \cite{DBLP:conf/bigdataconf/PreeceE19, leinweber2023Leveraging}.
In this example, operators can be mobility providers or public parties, e.g., states of a federal state or member states of a supranational union. This scenario also generalizes to similar transaction processing tasks for electric vehicle charging or smart grid scenarios~\cite{DBLP:journals/iotj/MollahZNLZGKY21} and, potentially, even to some Central Bank Digital Currency~\cite{raskin2018cdbc,ozili2023cdbc} scenarios. 
These use cases require high throughput, at least in the tens of thousands of transactions per second, high availability, and resilience.
Thus, we are interested in understanding how such a ``Not eXactly Byzantine'' (\nxb) fault model can be exploited by protocol design and, correspondingly, to evaluate gains and trade-offs.

We propose an approach called \nxbft{} that makes use of Trusted Execution Environments (TEEs) as well as asynchronous and graph-based atomic broadcast. TEEs are chosen since they are able to significantly increase efficiency and performance~\cite{DBLP:conf/eurosys/BehlDK17,DBLP:conf/ccs/WangDNRZ22,DBLP:conf/ipps/DecouchantKRY24}. Asynchronous, graph-based atomic broadcast is chosen since it has demonstrated resilience and impressive throughput \cite{danezis2022narwhal}.
\nxbft{} is based on TEE-Rider \cite{DBLP:conf/wdag/LeinweberH23},  an asynchronous and graph-based atomic broadcast protocol with a Byzantine fault tolerance of $n > 2f$, where $n$ denotes the number of replicas and $f$ the number of faulty nodes. TEE-Rider itself is based on DAG-Rider \cite{DBLP:conf/podc/KeidarKNS21}.
To increase \emph{resilience, throughput} and \emph{practicality}, we extend the \nxb{} fault model to an operating model:
While, in general, operating in a Byzantine and asynchronous environment, we assume that operators (1) do not attack their TEE themselves, (2) do not manipulate the business logic of their application, and (3) enable sufficiently long phases of synchrony required for ``maintenance work''.
The \nxb{} model allows maximum utilization of the inherent parallelism of TEE-Rider, thereby increasing throughput significantly, and to overcome fundamental issues with crash recovery.

The main contribution of this paper is to show that the \textit{combination} of the \nxb{} operating model and the \nxbft{} approach  is highly beneficial in terms of efficiency, crash resilience, and throughput.
In particular, our contributions are as follows:

\paragraph*{Co-Design of Assumptions and Algorithms}
With \nxbft, we designed a performance and resilience-oriented SMR protocol with crash recovery optimized for the \nxb{} model. 
\nxbft{} requires only half a network round trip for a logical round.
\nxbft{} uses a small TEE for non-equivocation and a common coin; the goal is to reduce costly cryptography and context switches.
We exploit the \nxb{} model such that each message exchanged between replicas at the same time distributes new client requests and drives the consensus process.

\paragraph*{Practicality in the Light of Fundamental Challenges}
We observed that the intricacies of the combination of TEE and asynchrony have not yet been fully explored. Recovery of a crashed replica $p_c$ requires resetting the TEE of $p_c$ and agreement between all correct replicas on the message history of $p_c$, i.e., agreement on a checkpoint.
Asynchrony makes it impossible for a replica to be sure that it has received the same of $p_c$'s messages as other replicas, denying quorum-based checkpoint establishment. 
We discuss these issues and a corresponding workaround relying on synchrony for liveness.
Additionally, we propose a pragmatic setup procedure.

\paragraph*{High Throughput and Resilience}
We implemented \nxbft{} and provide a throughput-latency trade-off analysis in comparison to Chained-Damysus \cite{DBLP:conf/eurosys/DecouchantKRY22} (rotating leader) and MinBFT \cite{veronese2013efficient} (static leader) for up to 40 replicas and network round trip latencies up to $150\ms$\footnote{All implementations are available at \url{https://blinded.for/review}}. 
\nxbft{} outperforms MinBFT and Damysus in terms of throughput in all settings.
For $n=40$, \nxbft{} achieves a throughput of $178\kops$ for datacenter deployments and $20\kops$ for world-wide deployments.
The throughput-oriented optimizations based on the \nxb{} model have a latency penalty that we deem reasonable.
When small latencies are required, MinBFT and Damysus are at an advantage, with Damysus benefiting from the \nxb{} model in terms of throughput for small deployments.
In contrast to MinBFT and Damysus, \nxbft's performance is almost not affected by actual crash faults.
Please note that a general comparison between TEE-based and non-TEE-based approaches is not a goal of this paper. 

The remainder of this paper is structured as follows.
We give an overview on related work on hybrid fault models, asynchronous consensus, and recovery procedures in Sec.~\ref{sec:background}.
In Sec.~\ref{sec:overview}, we define the \nxb{} model and give an overview on \nxbft.
We describe the \nxbft{} design in Sec.~\ref{sec:protocol} in detail.
Our experimental setup and empirical findings are presented in Sec.~\ref{sec:performance}.
We discuss the implications of our approach and future research directions in Sec.~\ref{sec:discussion} and conclude with Sec.~\ref{sec:conclusion}.

\section{Background and Related Work}
\label{sec:background}
Since \nxb{} is a variant of a hybrid fault model and NxBFT is constructed and optimized using atomic broadcast based on directed acyclic graphs (DAGs), we present corresponding related work. Furthermore, since recovery is particularly challenging when TEEs are used, we also address this specific aspect.
We assume the reader is familiar with the concepts of State Machine Replication (SMR)~\cite{DBLP:journals/csur/Schneider90} and of classical Byzantine fault tolerant SMR approaches like PBFT~\cite{DBLP:journals/tocs/CastroL02}.

\subsection{Hybrid Fault Models}

The core idea of hybrid fault models, which have been investigated for at least two decades \cite{DBLP:conf/srds/CorreiaNV04}, is to equip replicas with a trustworthy subsystem that is assumed to only fail by crashing and enforcing non-equivocation:
A replica cannot send two messages with different contents to different peers in the same context without being noticed \cite{DBLP:conf/sosp/ChunMSK07, DBLP:conf/sac/CorreiaVL10, DBLP:conf/podc/ClementJKR12, DBLP:conf/podc/MadsenD20}.
The non-equivocation property saves rounds of communication and allows operating BFT SMR with a fault tolerance of $n>2f$. 
With the advent of hardware-based Trusted Execution Environments (TEEs) like Intel SGX \cite{DBLP:journals/iacr/CostanD16}, researchers were able to propose practical systems using a trusted subsystem in the partially synchronous model adopting ideas from PBFT \cite{DBLP:conf/nsdi/LevinDLM09,veronese2013efficient,DBLP:conf/eurosys/BehlDK17,DBLP:journals/tc/LiuLKA19}.
These works rely on a TEE to provide confidential execution of code. 
Additionally, a TEE must also be able to remotely attest to the integrity of itself and code running on it.
The TEE implements a signature service that maintains an increment-only counter. 
When broadcasting a message, a replica has to attach a signature and a corresponding counter.
Receiving replicas will only accept one message for each possible counter value for each replica.
Under the assumption that the TEE will assign each counter value only once, receiving replicas can be sure that they all see the same message for the same counter value.
Recent work focuses on the rotating leader paradigm \cite{DBLP:conf/eurosys/DecouchantKRY22, DBLP:conf/ipps/DecouchantKRY24}, the asynchronous \cite{DBLP:journals/jsa/FuWSZ22,DBLP:conf/wdag/LeinweberH23,DBLP:journals/corr/abs-2501-01062} timing model, and alternative trusted subsystems~\cite{DBLP:conf/asplos/AguileraBGMXZ23}.
We show, in contrast to \cite{DBLP:conf/eurosys/GuptaRPCS23}'s claim, that trusted subsystems allow a highly parallelized and efficient consensus protocol design as it is also stated by \cite{DBLP:journals/corr/abs-2312-05714}.

\subsection{DAG-Based Atomic Broadcast}

Asynchronous consensus protocols promise increased resilience and throughput while at the same time being less complex than their partially synchronous counterparts \cite{DBLP:conf/ccs/MillerXCSS16,danezis2022narwhal}.
Hashgraph \cite{DBLP:conf/coins/BairdL20} and DAG-Rider \cite{DBLP:conf/podc/KeidarKNS21} were among the first asynchronous consensus protocols that use DAGs to encode the logical chronology of messages exchanged 
and derive the total order of transactions using a deterministic graph traversal.
In contrast to Hashgraph (and its TEE variant \cite{DBLP:journals/jsa/FuWSZ22}), DAG-Rider simplifies the consensus derivation and the message exchange by building the graph deterministically.
With Narwhal \cite{danezis2022narwhal}, the messages exchanged were reduced from $O(n^3)$ to $O(n^2)$ compared to DAG-Rider.
TEE-Rider \cite{DBLP:conf/wdag/LeinweberH23} showed that DAG-Rider can be compiled to withstand Byzantine faults with $n>2f$ replicas using a TEE-based signature service.
Ladelsky and Friedman~\cite{ladelsky2025quorum} generalize the proof of \cite{DBLP:conf/wdag/LeinweberH23} and analyze the impact of quorum sizes on termination properties.
Yanadmuri et al.~\cite{DBLP:conf/opodis/YandamuriANR22} showed how to reach quadratic communication complexity in the worst case using a small TEE. 
\nxbft{} builds upon TEE-Rider for increased throughput and for its less complex implementation; it further improves ideas from Narwhal to reach quadratic communication complexity.
In contrast to Fides \cite{DBLP:journals/corr/abs-2501-01062}, which pursues similar ideas to TEE-Rider, \nxbft{} improves both communication complexity and throughput and minimizes the size of the TEE-deployed code.
We comment on parallel work on leaderless atomic broadcast in \emph{partial} synchrony, in contrast to the work at hand, in Sec.~\ref{sec:discussion}.

\subsection{Crash Recovery and Reconfiguration}

When replicas experience unintended faults, e.g., in the case of hardware failures or simply when doing maintenance, they miss state updates and cannot participate anymore.
A recovery procedure allows a replica to catch up and be able to produce valid messages and to decide the validity of received messages~\cite{DBLP:journals/csur/Distler21}.
PBFT \cite{DBLP:journals/tocs/CastroL02} implements a checkpointing mechanism usable for recovery.
A recovery mechanism can be extended to support reconfiguration: replicas may join or leave the peer-to-peer network based on the ongoing consensus algorithm.
A TEE-aware recovery procedure requires the establishment of a new signature secret and it has to bring all correct replicas ``on the same page'' concerning the messages broadcast by the recovering replica.
To the best of our knowledge, CCF~\cite{DBLP:journals/pvldb/HowardAACCCDFJK23} and Achilles~\cite{niu2025achilles} are the only TEE-based SMR systems offering recovery.
Both are leader-based and operate in partial synchrony.
We show that such an approach is incompatible with the asynchronous nature of \nxbft's atomic broadcast.

\section{\texorpdfstring{\nxbft: Overview}{NxBFT: Overview}}
\label{sec:overview}

We explicate our assumptions of the \nxb{} operating model and provide an overview of the protocol structure of \nxbft{}.

\subsection{Assumptions: NxB Operating Model}
\label{subsec:om}

We consider a federation of $n$ predefined operators offering a common service using State Machine Replication (SMR) for an arbitrary number of clients.
Each operator operates a replica; $f < \frac{n}{2}$ replicas and all clients may behave Byzantine.
We assume replicas to execute the same non-manipulated state machine limiting operators to omission faults when communicating with clients.
Each replica is equipped with a Trusted Execution Environment (TEE) capable of running arbitrary code (called enclave) that offers confidential and integrity protected computing as well as remote attestation to verify the authenticity of the TEE and its enclave.
The TEE can confidentially and authentically export (partial) enclave state (``sealing'').
The TEE is assumed to only fail by crashing; its internal state is lost when crashing.
If not explicitly stated, code is not executed inside the TEE. 
Replicas and clients communicate via secure point-to-point links. 
Messages can be reordered and arbitrarily delayed but not dropped.
Setup and recovery require synchrony for liveness.

\subsection{\texorpdfstring{Protocol: \nxbft{}}{Protocol: NxBFT}}
\label{subsec:building_blocks}

\begin{figure}
    \centering
    \includegraphics[width=\columnwidth]{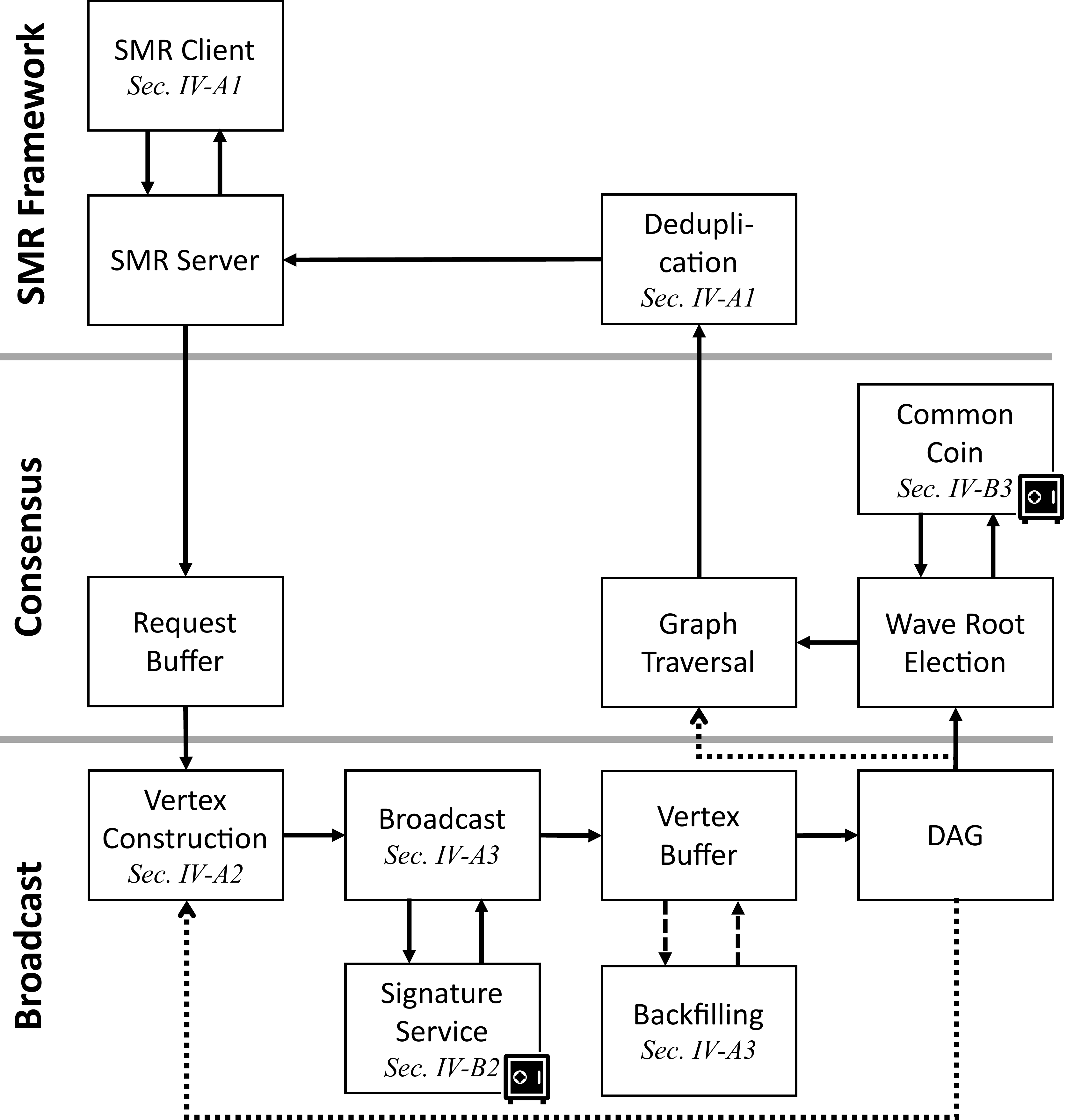}
    \caption{\nxbft{} components for normal operation (setup and recovery left out). The safe icon marks components executed inside the TEE. Solid arrows show the flow for a client request to be successfully ordered and executed. After being received by the replica, a request is buffered and eventually broadcast as a vertex payload using TEE-based signatures. A vertex is added to the DAG when its ancestry was already added (optional backfilling, dashed arrows). Each added vertex will eventually be ordered by a graph traversal; the corresponding root is selected using a common coin. After request deduplication, the request is output to the application layer and the response sent to the client. The dotted arrows highlight components using the DAG as input. Serif italic font names the section covering the component description.}
    \label{fig:nxbft}
\end{figure}

\nxbft{} makes use of the \nxb{} operating model: each replica uses its TEE for a signature service and a common coin. \nxbft{} builds on
TEE-Rider \cite{DBLP:conf/wdag/LeinweberH23} that maintains a DAG that is structured in rounds.
With the exception of setup and recovery, \nxbft{} operates asynchronously.
\nxbft{} is structured in three layers (see \Cref{fig:nxbft}): SMR framework, consensus, and broadcast. In the following, we describe a full Request--Broadcast--Consensus--Response cycle.

\paragraph{Request}
When a client wants to use the federated service, it composes a request containing the command and a sequence number and sends the request to a single replica (cf.~Sec.~\ref{subsubsec:client}).
A receiving replica buffers the request until it can be included in the payload of a broadcast round.

\paragraph{Broadcast}
In each round $r$ of the broadcast layer, each replica has to propose a new DAG vertex containing client requests to be ordered (cf.~Sec.~\ref{subsubsec:vertex}). 
Vertices $v$ are bound to a round $v.r$ and broadcast in a best effort fashion.
\nxbft{} uses a TEE-based signature service (cf.~Sec.~\ref{subsubsec:signature}) to prevent equivocation on the broadcast layer limiting faulty replicas to omission faults.
The counter of the signature service is used to enforce a FIFO ordering, i.e., a message with lower counter value must be delivered before a message with higher counter value can be delivered.
If received vertices contain an already processed counter value or an invalid signature, they are dropped.
A vertex needs to reference $\quorum$ vertices of the previous round as edges. 
Correct replicas will always reference their own vertex of the previous round as an edge.
Due to asynchrony, vertices for rounds $v.r < r$ or $v.r > r$ may be received any time.
Received valid vertices are buffered until they can be added to a replica's local DAG.
A vertex can be moved from buffer to DAG if the replica already transitioned to the vertex' round and the replica's DAG contains all referenced vertices. 
By transitively applying this requirement, the \emph{complete} ancestry of a newly added vertex must already be part of the DAG.
If vertices of the ancestry are missing, the replica will ask its peers to retransmit said vertices (backfilling).
DAG construction and backfilling-based reliable broadcast (cf.~Sec.\ref{subsubsec:broadcast}) in combination build a causal order broadcast \cite[Module 3.9]{DBLP:books/daglib/CachinGR2011}; the DAG encodes the causal order of exchanged messages.
A replica completes round $r$ and transitions to round $r+1$ when it added at least $\quorum$ valid vertices for round $r$ to the DAG.

\paragraph{Consensus}
Every disjoint four consecutive broadcast rounds are grouped into a \emph{wave}:
round $r$ belongs to wave $w=\lceil \frac{r}{4} \rceil$.
When a wave $w$ is completed, that is, its fourth round is completed, the consensus layer is invoked:
It tries to derive agreement on a total order of requests of wave numbers smaller than $w$ for all correct replicas. The wave construction works as follows. Assume a wave $w$ is complete.
A vertex of the first round of wave $w$ is selected as wave root; since the TEE-based common coin (cf.~Sec.~\ref{subsubsec:coin}) is used for selection, all correct nodes select the same vertex.
If the wave root is part of a replica's local DAG when it is selected and at least $\quorum$ vertices of $w$'s fourth round have a path to it (``direct commit rule''), the replica can commit the wave:
The wave root is then used as starting point for a pre-defined deterministic graph traversal to order all not yet ordered requests of previous waves.
If the wave root is not part of the DAG, the wave cannot be committed yet.
A replica will check during direct commits of future waves if it can retrospectively commit a wave.
For a retrospective commit, a single path between old and new wave root is sufficient.
The construction of a wave ensures that the root of the next wave (and all future waves) has a path to a previous root. For proofs of correctness, we refer the reader to \cite[Lemma 3]{DBLP:conf/wdag/LeinweberH23} and \cite[Proposition 2]{DBLP:conf/podc/KeidarKNS21}.
However, it is important to note that the consensus layer works solely on the local DAG without any additional communication.

\paragraph{Response}

The requests are executed by the server-side application in the order they were added to the vertex by the proposing replica.
The sequence number of a request is used for deduplication (cf.~Sec.~\ref{subsubsec:client}): the combination of client id and sequence number is output exactly once to the application.
After execution, the result is sent as a response to the initially requesting client.

\section{\texorpdfstring{\nxbft: Design Considerations}{NxBFT: Design Considerations}}
\label{sec:protocol}

\begin{figure*}[t!]
    \centering
    \subfigure[TEE-Rider: $n^2 + n^3$\label{fig:comm:rider}]{
    \includegraphics[height=2.5cm]{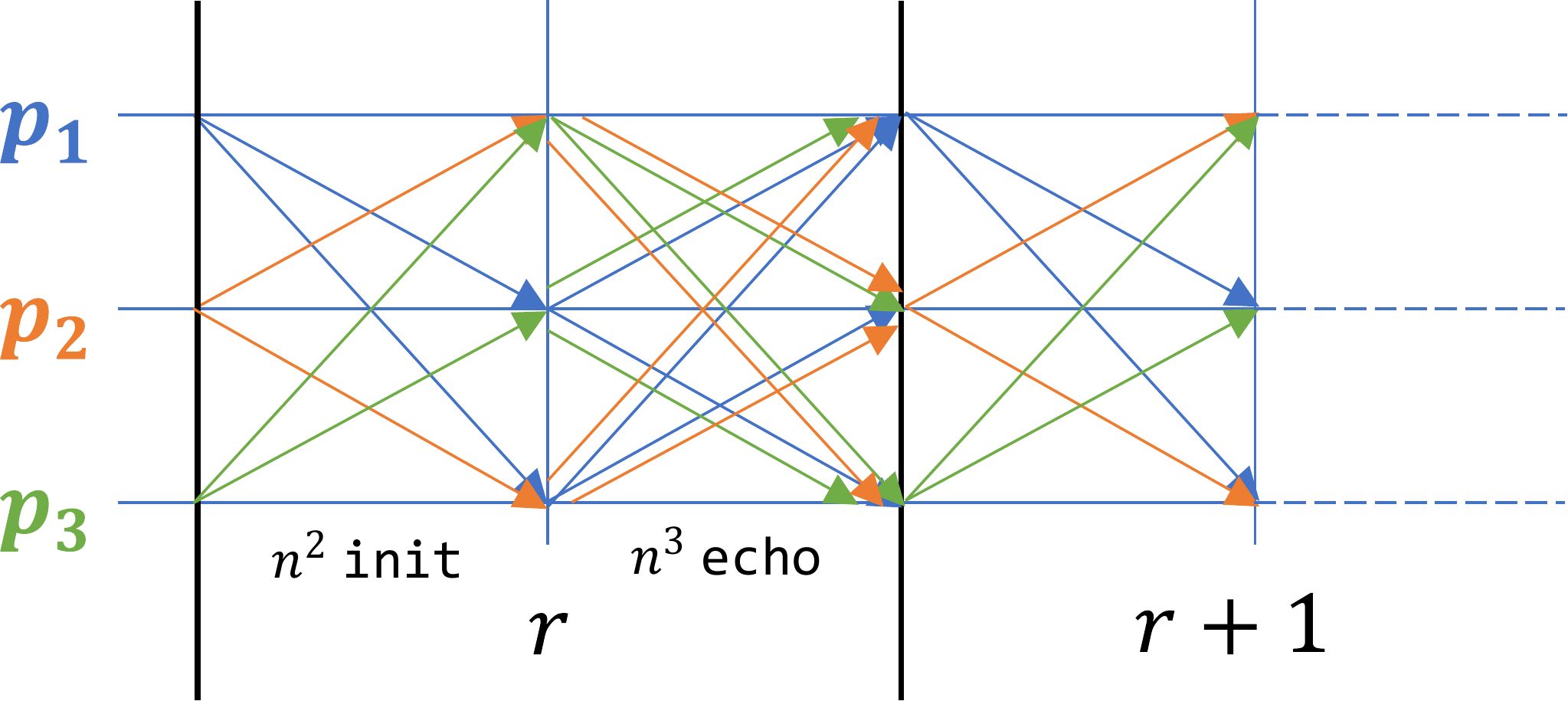}
    }
    \subfigure[Narwhal: $3n^2$\label{fig:comm:narwhal}]{
        \includegraphics[height=2.5cm]{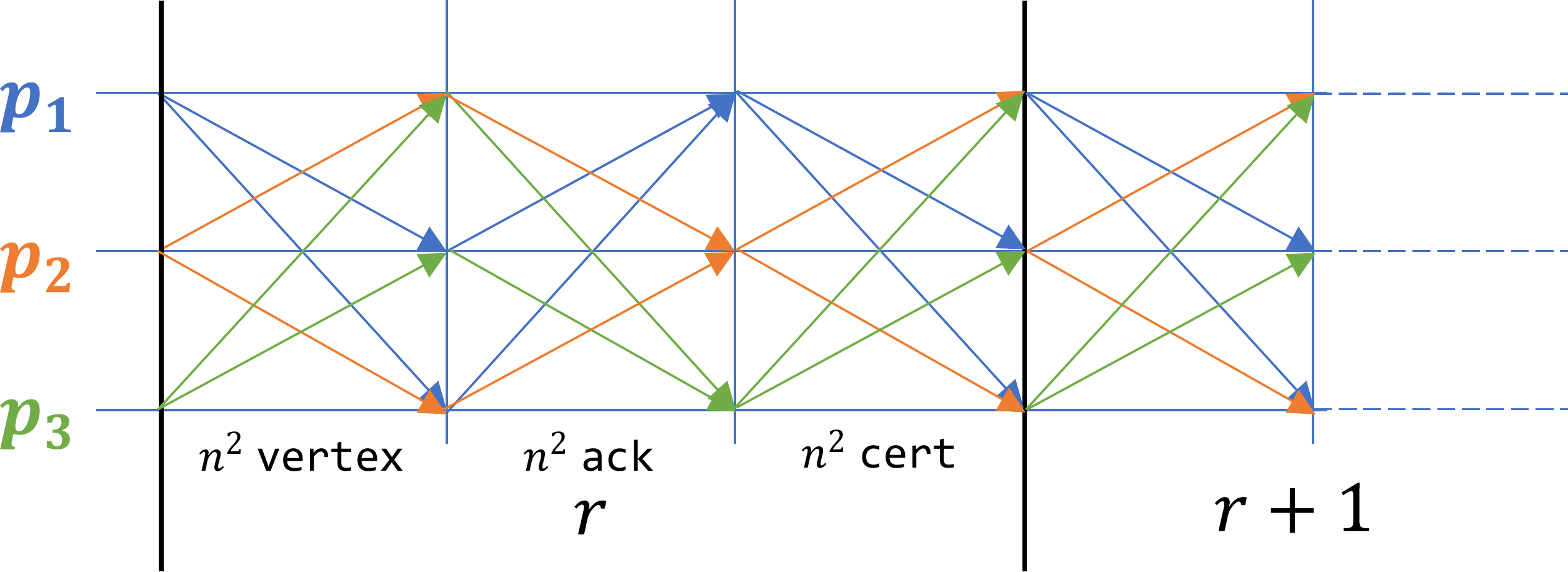}
    }
    \subfigure[\nxbft: $n^2$\label{fig:comm:nxbft}]{
        \includegraphics[height=2.5cm]{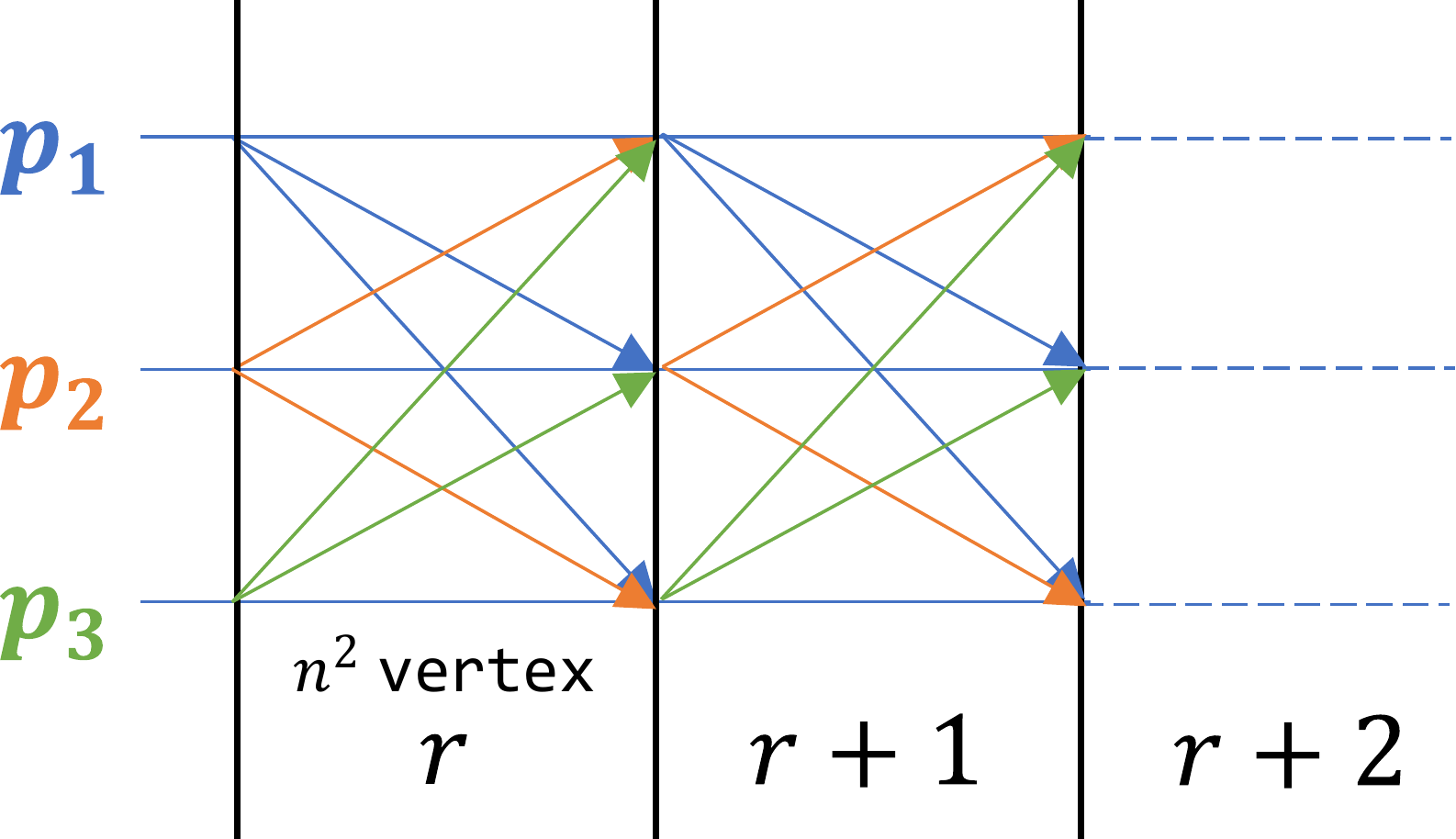}
    }

    \hspace{0.2cm}
    \caption{Common case communication patterns of TEE-Rider \cite{DBLP:conf/wdag/LeinweberH23}, Narwhal \cite{danezis2022narwhal} and \nxbft{} (this work). Blue vertical lines delimit communication rounds, black vertical lines delimit rounds of the broadcast layer. The colors of the arrows signify the payloads' originator. \nxbft{} proceeds with one round of communication per broadcast round.}
    \label{fig:communication}
\end{figure*}

We discuss various design considerations in detail and
refer the reader to Appendix~\ref{sec:correctness} for arguments of correctness.

\subsection{Increasing Throughput}
\label{subsec:scaling}

\subsubsection{\texorpdfstring{\nxb{} Client Model}{NxB Client Model}}
\label{subsubsec:client}
Clients select a replica at random and unicast their request accompanied with a client id and a sequence number.
Once the consensus layer decided, all correct replicas will input the request to the state machine and the output of the (application logic) state machine is sent as a response.
In the \nxb{} model, a single response suffices for the client to complete the request.
This mechanism allows each replica to propose a disjoint set of requests in each round scaling up throughput with $n$.

When the client issues its request, it starts a timer.
If the selected replica does not answer within the time interval, the client selects a different replica for its request (and so forth).
The client is guaranteed that its request will eventually be ordered, however, a request can appear in the DAG up to $n$ times.
\nxbft{} employs a simple deduplication logic: a correct client will only have one unanswered request at a time.
The replicas store for each client the client's last executed sequence number.
A request is only input to the state machine when its sequence number is greater than the stored sequence number.
If a client issues more than one request simultaneously or equivocates on sequence numbers, it is possible that some of its requests will not be answered.

\subsubsection{Vertex Construction}
\label{subsubsec:vertex}
Asynchronous algorithms typically rely on valid messages from peers to make progress (see, e.g., \cite[Algorithm 2]{DBLP:conf/podc/KeidarKNS21}).
In the case of TEE-Rider, a replica needs at least $\quorum$ valid vertices for its current round to transition to the next round and to be allowed to broadcast its next vertex.
To increase the achievable throughput, a replica holds back its vertex until it can propose at least a configurable amount of client request as the vertex' payload.
If now only a few clients issue requests, the system would come to a halt as the quorum for a round to complete would need a lot of time to be established.
The \nxb{} client model worsens this problem as requests are equally distributed among the replicas. 
We address this issue by working with (local) timers on the side of the replica:
When a replica is not able to broadcast a vertex containing enough client request within a pre-defined time interval, the replica will broadcast the vertex anyway.
This allows other replicas to complete the round.

\subsubsection{Backfilling-Based Reliable Broadcast}
\label{subsubsec:broadcast}
Inspired by Narwhal \cite{danezis2022narwhal} and in contrast to TEE-Rider \cite{DBLP:conf/wdag/LeinweberH23}, correct replicas do not echo vertices by default.
Instead, if an ancestor is unknown, i.e., it was neither already added to the DAG nor buffered, a request for the missing vertex is sent to all replicas.
Due to asynchrony, correct replicas cache vertex requests if they cannot answer them directly: It is possible that the corresponding vertex arrives delayed and the replica receiving the vertex request is the only correct replica that receives the requested vertex due to faults by the sender.
A different option would be for replicas to repeat their request after some time trading potentially unbounded memory consumption with algorithmic complexity.
TEE-Rider uses a proactive single echo broadcast leading to $n^2 + n^3$ messages per broadcast round (\Cref{fig:comm:rider}).
Narwhal uses a three-way handshake to produce quorum certificates that certify the availability of vertices.
\nxbft{} can save a constant communication factor as we do not rely on quorum certificates:
In the common case, Narwhal needs $3n^2$ messages per broadcast round (\Cref{fig:comm:narwhal}), whereas \nxbft{} proceeds with one $n^2$ communication step (\Cref{fig:comm:nxbft}).
In the worst case, both send up to $n$ backfilling requests to all replicas and receive $n$ replies, leading to $5n^2$ messages for Narwhal vs. $3n^2$ messages for \nxbft{} per broadcast round.

\subsection{Small Enclave for Efficiency and Resilience}
\label{subsec:enclave}

The goal of our enclave design is a small trusted computing base and the reduction of costly context switches for efficient and resilient execution.
The enclave of \nxbft{} combines the well-known concept of a TEE-based signature service to prevent equivocation \cite{DBLP:conf/nsdi/LevinDLM09, veronese2013efficient} with a common coin based on a cryptographically secure pseudorandom number generator (PRNG) and a pragmatic setup procedure. 
To be able to toss a coin, a replica has to convince the coin implementation that it made sufficient progress on the consensus layer by providing sufficient signature service signatures from peering replicas.
In the following, we use the term (public/secret) enclave keys to refer to the keys used for the signature service in contrast to long-living (public/secret) replica keys.

\subsubsection{Setup}

During setup, \nxbft{} enclaves mutually attest and verify their integrity, reach consensus on the deployed enclave code as well as the peers' public enclave keys to prevent simulation attacks, and establish the authenticated channels between replicas.
The second objective of the setup protocol is to initialize the PRNG underlying the common coin with a collaboratively generated random seed.
Due to the required participation of all $n$ replicas in both of the above tasks, \nxbft{} cannot tolerate but reliably detect faults during setup.
We construct the setup protocol using $n$ synchronous authenticated reliable broadcast instances with a fault tolerance of $n>f$~\cite{DBLP:journals/jacm/PeaseSL80} to preempt equivocation by faulty replicas.
We describe the setup protocol of \nxbft{} as a one-way handshake between replicas $p_i$ and $p_j$.
A replica $p_i$ also performs this setup handshake with itself.

Upon initialization, the enclave of $p_i$ initializes its state variables and generates both a new enclave key pair and a random seed share for the later initialization of the common coin PRNG.
Replica $p_i$ now uses a synchronous authenticated single-echo reliable broadcast to disseminate its identifier and its signed attestation certificate carrying its public enclave key.
Replica $p_j$ waits to receive $n$ validly signed echo messages carrying consistent and valid attestation certificates.
Replica $p_j$ then sends its encrypted seed share $p_i$ (using $p_i$'s public enclave key), thereby completing its part of the handshake.
Replica $p_i$ then invokes its enclave with the received encrypted seed share. The enclave decrypts the seed share and XOR's it with the current seed value.
Once a replica has completed all $n$ handshakes, it has completed the setup protocol and broadcasts a ready message to all replicas.
Replicas start a timer for each handshake, the expiration of which raises an error and leads to an abortion of the entire setup.
Similarly, conflicting or invalid messages and certificates raise errors and lead to an abort.

\subsubsection{Signature Service}
\label{subsubsec:signature}
The signature service is used to enforce non-equivocation on messages.
Each signature is accompanied with a unique counter value.
Receiving replicas accept only one message per counter value.
On initialization, each enclave generates an asymmetric key pair used exclusively for the signature service and sets its counter value $c$ to 0.
Attestation certificates generated by an enclave contain its public enclave key and are signed with a replica key during setup and recovery in order to establish a binding between these two key pairs.
Thus, enclaves learn the public enclave keys of each other when verifying attestations ensuring the key stems from a running enclave.
When a replica requests a signature for a message, the enclave signs the combination of message and current counter value with its secret enclave key, increments its counter by one, and returns signature and corresponding counter.
Attestation certificates and signatures can be verified outside of enclaved execution as well.

\subsubsection{Common Coin}
\label{subsubsec:coin}
Asynchronous consensus protocols rely on randomness to circumvent the FLP impossibility \cite{DBLP:conf/pods/FischerLP83,DBLP:conf/podc/Ben-Or83}.
\nxbft{} uses the common coin to select a wave root, i.e., it tosses a number in $\{0, \ldots , n-1\} \subset \mathbb{N}_0$.
The common coin must be fair, it must produce the same output for every replica and wave, it must not be predictable until at least $\quorum$ replicas tossed for a wave, and it has to be revealed when at least $\quorum$ replicas toss for a wave.
Typically, common coins are implemented using threshold signatures \cite{DBLP:conf/pkc/Boldyreva03, DBLP:journals/corr/abs-2502-03247}.
We argue that, when already using a TEE, it seems rational to avoid the use of rather expensive cryptographic computations.
The random bit string established by XOR'ing the local seed shares during setup is used to seed a PRNG.
The coin implementation allows a coin toss whenever it can be convinced that four rounds of \nxbft's broadcast layer were successfully completed by the replica.
Vertices have to be from the last round of the wave that is being tossed for.
The implementation maintains a variable that is initially set to four and resembles the expected round in which a coin toss is allowed.
A replica has to provide at least $\quorum$ validly signed vertices of different replicas and the expected round, proving the wave was completed by at least one correct replica.
If the replica supplied sound evidence, the expected round is incremented by four and the next value of the PRNG returned.

\subsection{Limits of and Approach to Crash Recovery}
\label{subsec:recovery}

Faults like hardware failures, human configuration errors, and maintenance require that a replica can be put back into the condition to validate received messages and produce valid messages.
We are interested in an algorithmic solution that can recover replicas while allowing the system to continuously operate.
Such a recovery procedure must ensure that the recovery does not allow (1) equivocation and (2) to learn coin values beforehand. For \nxbft{} in particular, the recovery procedure has to be aware of the TEE and its state. 
We first present two challenges one is facing in the design of a recovery procedure for \nxbft.
Then we present a recovery procedure that is Byzantine safe under asynchrony but requires participation of \textit{all} replicas for liveness.

\subsubsection{Impossibility Argument}

\nxbft's only \emph{active} agreement primitive is an asynchronous TEE-based reliable broadcast with a fault tolerance of $n>f$. 
From the reliable broadcast message history and state, all decisions are passively derived.
Thus, recovering a \nxbft{} replica is, for the most part, recovering the reliable broadcast module leading to two issues:  (1) the recovery procedure cannot use the ongoing asynchronous consensus for coordination and (2) it is impossible to use a quorum-based decision for recovery.

We explain the first impossibility with the following example.
Assume three replicas Alice, Bob, and Charlie with Charlie trying to recover and broadcasting a specially crafted request containing its new public enclave key authenticated with Charlie's public replica key.
Alice and Bob both receive Charlie's request, propose it for ordering, and eventually reach consensus.
Due to asynchrony, after reaching consensus on the recovery, Alice receives a vertex from Charlie signed with the old enclave key.
Consequently, she will reject it. 
Bob, however, received the same vertex before reaching consensus with Alice and accepted it.
If both now sort their graphs, they will get different orderings breaking the consensus layer's safety.
Even when requesting the vertex from Bob because he references it in his vertex, Alice has no possibility to judge on Bob's and Charlie's honesty. Thus, if multiple identities of a peer's enclave exist simultaneously, equivocation can happen.

Recovery requires all correct replicas to reach agreement on the valid vertices of the recovering replica.
Additionally, to prevent the necessity for rollbacks, the input of all correct replicas to such a decision has to be honored.
This forbids a quorum-based approach, as any quorum reached in asynchrony may outvote a correct replica:
In a TEE-based reliable broadcast with a fault tolerance of $n>f$, a receiving replica can, on successful signature verification, deliver immediately. 
Therefore, a correct replica Alice at an arbitrary but fixed point in time may be the only correct replica having delivered a certain value, e.g., of Bob.
This is not in conflict with the totality property as Alice will relay the value and eventually all correct replicas will deliver as well. 
If Bob now crashes and recovers, all correct replicas need to agree on what Bob sent and what was potentially delivered.
Any quorum smaller than $n$ has a chance to outvote Alice: due to asynchrony, there is neither a guarantee that all votes building the quorum are cast by correct replicas nor that the votes used for a quorum are the same for all correct replicas.
Thus, there exist traces in which Alice would need to ``undeliver'' Bob's value breaking the reliable broadcast totality property \cite[Module 3.12]{DBLP:books/daglib/CachinGR2011}.
If now any precondition is lifted, i.e., asynchrony, prevention of rollbacks, or TEEs preventing equivocation, this impossibility does not hold anymore. 

\subsubsection{Recovery Protocol}

The recovery protocol is a variation of classical interactive consistency \cite{DBLP:journals/jacm/PeaseSL80}.
The recovery protocol assumes correct operators to create a backup of state and enclave at least after setup and after every successful recovery.
The enclave backup uses sealing and contains all state except the secret enclave key and the counter value.
To circumvent the introduced impossibilities, the protocol requires the input of \emph{all} replicas to reach a recovery decision. 
Based on the backed-up state, a recovering replica $p_c$ requests a recovery consensus of all replicas by broadcasting a \emph{RecoveryRequest} message.
Receiving replicas initialize a consensus procedure to achieve agreement among all $n$ replicas on $p_c$'s message history.

We assume that at any time, a number of $r < \quorum$ replicas request to recover.
First, the recovering replica $p_c$ will try to recover the enclave by providing an encrypted enclave state export.
The replica $p_c$ broadcasts a \emph{RecoveryRequest} with the new attestation certificate including $p_c$'s new public enclave key.
A receiving replica $p_i$ will verify the attestation, delete all buffered vertices that belong to $p_c$, and identify all vertices of $p_c$ in its graph, i.e., $p_c$'s message history.
Finally, $p_i$ broadcasts a \emph{RecoveryProposal} for $p_c$ with $p_c$'s new attestation certificate and the message history as payload.
As soon as $n$ RecoveryProposals with equal attestation certificates are accepted, a correct replica $p_i$ will identify the proposal with the longest valid vertex chain and broadcast it along with $p_c$'s new attestation certificate as a \emph{RecoveryCommit}.
Each replica signs a RecoveryCommit with their secret replica key.
On the receipt of $\quorum$ consistent RecoveryCommits, a replica $p_i$ is able to complete the recovery procedure.
Replica $p_i$ will add all unknown vertices from the commit to the vertex buffer and set the expected counter value for $p_c$ to $0$.
Additionally, $p_i$ will provide the new attestation certificate of $p_c$ and the signatures from the collected RecoveryCommits to its enclave that, on successful verification, will exchange the public enclave key of $p_c$; $p_i$ can now process new messages of $p_c$.
The recovering replica $p_c$ will derive the first round it is allowed to propose a new vertex for from the RecoveryCommits; missed vertices are fetched using the backfilling mechanism.

\section{Practical Evaluation}
\label{sec:performance}

We investigate the impact of the \nxb{} client model, scaling behavior of \nxbft{} with varying payload sizes and network conditions, and the behavior under faults in comparison to MinBFT~\cite[Sec. 4]{veronese2013efficient} and Chained-Damysus~\cite[Sec. 7]{DBLP:conf/eurosys/DecouchantKRY22}. We also conduct measurements of the recovery procedure.

\subsection{Experiment Setup and Implementation}

We use the ABCperf framework \cite{DBLP:conf/middleware/SpannagelLCH23} as the basis for our experiments.
ABCperf orchestrates the experiment, implements the full communication stack, handles performance measurement, and emulates client behavior. 
ABCperf ships a no-op application with random byte payloads and a MinBFT implementation; we add \nxbft{} and Chained-Damysus (own implementation based on \cite[Sec. 7]{DBLP:conf/eurosys/DecouchantKRY22}) as atomic broadcast modules.
We run our experiments on a cluster of 25 servers of which we use one as an orchestrator, 20 for replicas (all Intel E-2288G, $64\,\mathrm{GB}$ main memory) and four for client emulation (AMD EPYC 9274F, $128\,\mathrm{GB}$ main memory) connected with $10\,\mathrm{GBit}$ interfaces.
ABCperf can emulate an increased network round trip latency.
ABCperf emulates $50\%$ of the configured latency in each direction.
The base network round trip latency of our cluster is $\sim0.15\ms$.
The actual number of replicas is equally distributed among the 20 replica hosts.

A single experiment run has the following pattern: the replicas are setup and perform the setup phase of the atomic broadcast algorithm. 
Then, we start to invoke client requests following a configurable constant frequency (``request rate'').
A client has at most one open request.
Now, whenever a request is to be made, the client emulator checks if it has an idle client, i.e., a client with no pending request, and issues the request according to the \nxb{} (random selection with fallback, one valid answer suffices) or BFT (broadcast, $\quorum$ valid answers required) client models. 
To prevent crashes of the emulation and experiment framework due to resource exhaustion, we have to limit the overall number of clients to two million.
After a pre-defined warm-up phase, the measurement phase starts.
We measure the end-to-end latency of each request as the elapsed time, observed by a client, between sending a request and getting a valid answer.
The achieved throughput is the number of successfully completed requests per second.

To ease implementation and, foremost, testing, the atomic broadcast state machines do not use any form of parallelism.
We deem this to be reasonable as we are not interested in actual numbers that can be expected for production deployments but the relative behavior of the three algorithms at investigation.
ABCperf's communication stack and client emulation, however, are highly parallelized.
All enclave code is executed with Intel SGX \cite{DBLP:journals/iacr/CostanD16}. 
Although we are aware of the limitations and weaknesses of SGX (e.g., \cite{DBLP:conf/uss/BulckMWGKPSWYS18}), in our experiments SGX serves as a viable way to account for the overhead of enclaved execution.
To ensure a correct implementation of all algorithms, we employ a combination of unit and randomized integration tests. 

The algorithm parameters that are fixed for all experiments are as follows and brought the best performance in a manual sensitivity analysis.
MinBFT requires a block to contain at least $10\mathrm{k}$ requests, Chained-Damysus and \nxbft{} require $100$ requests. 
All algorithms propose at least every $0.1\,\mathrm{s}$ a new block. 
MinBFT uses an exponentially increasing timer for timeouts.
Chained-Damysus uses the exponential increase and linear decrease as described in the paper \cite[Sec. 3]{DBLP:conf/eurosys/DecouchantKRY22}.
MinBFT and Chained-Damysus have an initial view timeout value of $3\,\mathrm{s}$.
The fallback timeout of an \nxb{} client is $5\,\mathrm{s}$.

\subsection{\texorpdfstring{\nxb{} Client Model Impact and Throughput Scaling}{NxB Client Model Impact and Throughput Scaling}}
\label{subsec:nxbpenalty}

To evaluate the impact of the client model and the algorithms' scaling behavior, we run \nxbft{}, Chained-Da\-my\-sus, and Min\-BFT with request rates between $50\kops$ and $800\kops$ (step size: $50\kops$) for $n\in \{ 3, 10, 20, 40 \}$ replicas with the BFT and the \nxb{} client model with eight repetitions of $60\s$ each.
To prevent measuring the effects of network speed and cryptographic operations and, instead, investigate the protocols' overhead, the requests have a zero byte payload and no additional network latency is emulated.


\begin{table}[t!]
\renewcommand{\arraystretch}{1.1}
\caption{Maximum Sustained Throughput (step size $50\kops$)\\in Dependence on Client Model\\($0\,\mathrm{B}$ Payload, $0.15\ms$ Network Round Trip Latency)}
\label{tab:peak_client}
\centering
\begin{tabular}{l||c|c||c|c||c|c}
  & \multicolumn{2}{c||}{\textbf{MinBFT}}& \multicolumn{2}{c||}{\textbf{Damysus}} & \multicolumn{2}{c}{\textbf{\nxbft}}  \\ 
       & BFT   & \nxb  & BFT   & \nxb  & BFT   & \nxb  \\ 
\hline
$n=3$  & $150$ & $150$ & $100$ & $350$ & $150$ & $350$ \\ 
$n=10$ & $100$ & $100$ & $100$ & $600$ & $150$ & $700$ \\ 
$n=20$ & $50$  & $50$  & $50$  & $150$ & $100$ & $450$ \\ 
$n=40$ & $50$  & $50$  & --    & $50$  & $50$  & $400$  
\end{tabular}
\end{table}

\Cref{tab:peak_client} shows the maximum request rates for which a stable system state was observed.
Using the \nxb{} client model, MinBFT achieves its peak performance for $n=3$.
Chained-Damysus' and \nxbft's throughput peaks at $n=10$.
When increasing $n$ to 20, the throughput of Chained-Damysus decreases by a factor of $\sim 4$. 
The throughput of \nxbft{} drops by a factor of $\sim 1.5$.
The table shows the maximum request rates for the BFT client as well.
MinBFT achieves the same throughput for both client variants.
Chained-Damysus' and \nxbft's throughput is at least halved when using the BFT client model. 
For $n=40$, Chained-Damysus did not stabilize for a request rate of $50\kops$.

\begin{figure}[t!]
    \centering
    \includegraphics[width=\columnwidth]{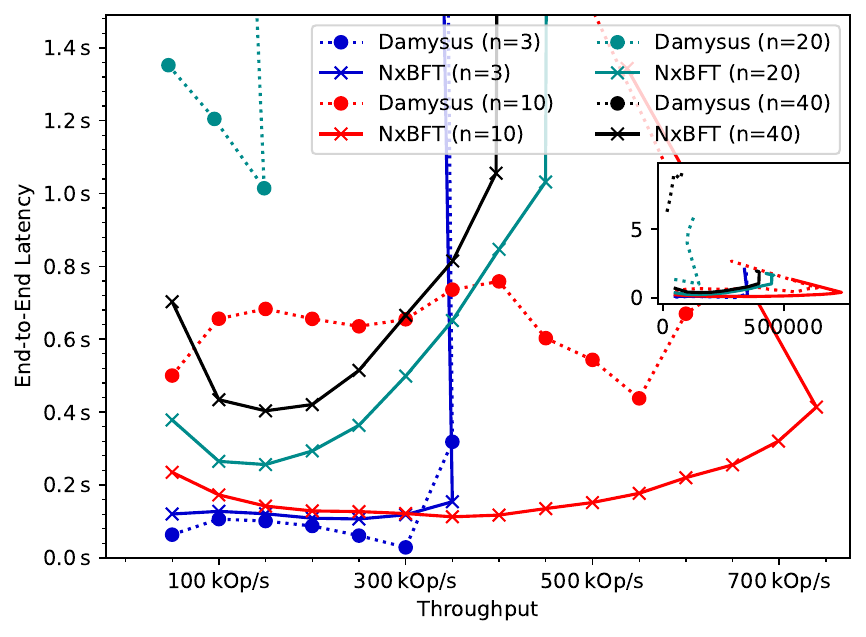}
    \caption{End-to-end-latency of Chained-Damysus (dotted lines) and \nxbft{} (solid lines) for request rates between $50\kops$ and $800\kops$ using the \nxb{} client model; each data point shown is an average of eight runs. The plot is limited on the y-axis; the full data is shown in the small window on the right. \Cref{tab:peak_client} indicates the corresponding maximum sustained request rates.}
    \label{fig:exp:scaling}
\end{figure}

\Cref{fig:exp:scaling} shows the corresponding end-to-end latencies for the \nxb{} client model.
Before saturation, MinBFT achieves latencies of $0.1\s$ for $n<20$ and $2.5\s$ for $n\geq 20$ (omitted for readability).
For \nxbft{} and Chained-Damysus, the latency increases when increasing the number $n$ of replicas. 
\nxbft{} achieves the best latency for $n=3$ and a request rate of $50\kops$ with $\sim 0.12\s$ and stays for all configurations, before saturating, below $1\s$. 
Chained-Damysus achieves the best latency for $n=3$ and a request rate of $350\kops$ with $\sim 0.01\s$.
For $n\geq20$, Chained-Damysus does not achieve latencies faster than $1\s$.

\begin{figure}[t!]
    \centering
    \includegraphics[width=\columnwidth]{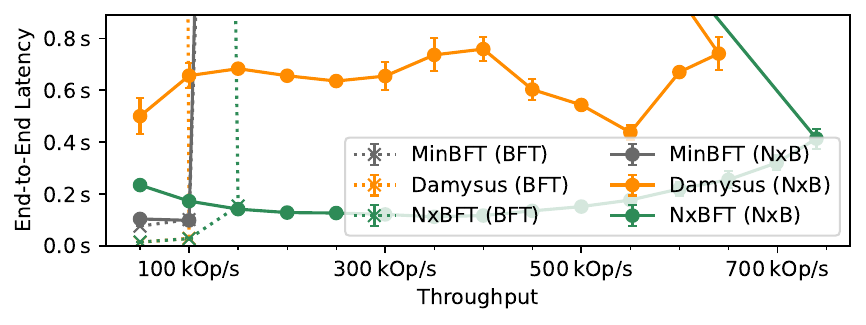}
    \caption{Investigation of the \nxb{} client latency penalty: end-to-end-latency of MinBFT, Chained-Damysus, and \nxbft{} for $n=10$ and request rates between $50\kops$ and $800\kops$ using \nxb{} (solid lines) and BFT (dotted lines) clients; each data point shown is an average of eight runs. The error bars indicate the $95\%$ confidence interval. The plot is limited on the y-axis.}
    \label{fig:exp:byzantine}
\end{figure}

\Cref{fig:exp:byzantine} compares the achieved end-to-end latencies of the two client models under investigation.
To increase readability, we select $n=10$.
Chained-Damysus and \nxbft{} achieve a speedup when using the BFT client model (factor $\sim 24$ for Chained-Damysus and factor $\sim 6$ for \nxbft).
MinBFT does not benefit from using the BFT client in terms of latency.

In summary, we can observe that the \nxb{} client model allows for load balancing when having a rotating leader or no leader at all but has a latency penalty.
While in MinBFT the current leader has to handle \emph{every} client connection, Chained-Damysus and \nxbft{} replicas handle, in expectation, $\frac{1}{n}$th of all client connections.
Additionally, \nxbft{} shows the benefit of each message exchanged carrying client requests: while the workload in terms of messages exchanged and, thus, cryptographic operations performed increases when increasing~$n$, \nxbft{} outperforms Chained-Damysus in terms of throughput.
The reduced connection load allows Chained-Damysus to scale as well, but for $n>10$ the increasing work to be done by the atomic broadcast eliminates those benefits without producing as much value as it is the case for \nxbft.
However, due to the random selection of a replica by a \nxb{} client, requests have to wait longer for block inclusion. 
Especially Chained-Damysus suffers significantly, as, in the worst case, a request has to wait $2n$ blocks for inclusion since Chained-Damysus requires a replica to be in charge for two blocks \cite[App. B]{Decouchant2022DamysusExtended}.
This maximum waiting time is reached when the client selected the replica which just proposed its second block.
For MinBFT, the client choice has no impact as the static leader has to handle every request anyways.

We observe that the servers executing the replica code have on average $\sim25\%$ total CPU usage but the process executing the atomic broadcast logic utilizes one CPU core permanently to the maximum.
Suitable parallelization at each replica \cite{DBLP:conf/eurosys/BehlDK17,danezis2022narwhal,einars2024enabling} may improve scaling behavior even further.

\subsection{Impact of Network Properties and Payload}

\begin{table*}[t!]
\renewcommand{\arraystretch}{1.9}
\caption{Maximum Sustained Throughput (in $\kops$, bold) and Corresponding Intermediate Decision Times (in $\ms$)\\Depending on Network Size and Latency with $256\,\mathrm{B}$ Payload Size} 
\label{tab:network_peak}
\centering
\begin{tabular}{l||c|c|c|c||c|c|c|c||c|c|c|c}
\multirow{2}{*}{\makecell{Network\\Round Trip\\Latency}}  & \multicolumn{4}{c||}{\textbf{MinBFT}}& \multicolumn{4}{c||}{\textbf{Chained-Damysus}} & \multicolumn{4}{c}{\textbf{\nxbft}}  \\ 
  & $n=3$  & $n=10$ & $n=20$   & $n=40$  & $n=3$   & $n=10$   & $n=20$   & $n=40$   & $n=3$   & $n=10$  & $n=20$  & $n=40$  \\ 
\hline
$0.15\ms$ & \makecell{ $\mathbf{69}$ \\ $124$ } & \makecell{ $\mathbf{28}$ \\ $253$ } & \makecell{ $\mathbf{18}$ \\ $129$ } & \makecell{ $\mathbf{11}$ \\ $137$ } & \makecell{ $\mathbf{223}$ \\ $7$ } & \makecell{ $\mathbf{132}$ \\ $41$ } & \makecell{ $\mathbf{64}$ \\ $65$ } & \makecell{ $\mathbf{36}$ \\ $46$ } & \makecell{ $\mathbf{439}$ \\ $24$ } & \makecell{ $\mathbf{584}$ \\ $23$ } & \makecell{ $\mathbf{268}$ \\ $109$ } & \makecell{ $\mathbf{178}$ \\ $396$ } \\
$5\ms$ & \makecell{ $\mathbf{63}$ \\ $121$ } & \makecell{ $\mathbf{25}$ \\ $184$ } & \makecell{ $\mathbf{17}$ \\ $143$ } & \makecell{ $\mathbf{11}$ \\ $149$ } & \makecell{ $\mathbf{120}$ \\ $28$ } & \makecell{ $\mathbf{64}$ \\ $69$ } & \makecell{ $\mathbf{56}$ \\ $177$ } & \makecell{ $\mathbf{31}$ \\ $123$ } & \makecell{ $\mathbf{304}$ \\ $69$ } & \makecell{ $\mathbf{234}$ \\ $247$ } & \makecell{ $\mathbf{191}$ \\ $269$ } & \makecell{ $\mathbf{99}$ \\ $486$ } \\
$35\ms$ & \makecell{ $\mathbf{49}$ \\ $120$ } & \makecell{ $\mathbf{20}$ \\ $121$ } & \makecell{ $\mathbf{13}$ \\ $126$ } & \makecell{ $\mathbf{11}$ \\ $153$ } & \makecell{ $\mathbf{29}$ \\ $150$ } & \makecell{ $\mathbf{20}$ \\ $791$ } & \makecell{ $\mathbf{14}$ \\ $539$ } & \makecell{ $\mathbf{3}$ \\ $323$ } & \makecell{ $\mathbf{91}$ \\ $352$ } & \makecell{ $\mathbf{55}$ \\ $680$ } & \makecell{ $\mathbf{45}$ \\ $1\,180$ } & \makecell{ $\mathbf{28}$ \\ $1\,404$ } \\
$150\ms$ & \makecell{ $\mathbf{33}$ \\ $111$ } & \makecell{ $\mathbf{15}$ \\ $117$ } & \makecell{ $\mathbf{12}$ \\ $123$ } & \makecell{ $\mathbf{9}$ \\ $137$ } & \makecell{ $\mathbf{6}$ \\ $1\,641$ } & \makecell{ $\mathbf{4}$ \\ $1\,580$ } & \makecell{ $\mathbf{1.2}$ \\ $1\,218$ } & \makecell{ $\mathbf{0.2}$ \\ $766$ } & \makecell{ $\mathbf{45}$ \\ $1\,299$ } & \makecell{ $\mathbf{26}$ \\ $1\,924$ } & \makecell{ $\mathbf{24}$ \\ $1\,625$ } & \makecell{ $\mathbf{20}$ \\ $1\,680$ } \\
\end{tabular}
\end{table*}

Graph-based and leader-rotating protocols are known for being highly dependent on the network latency between replicas.
We investigate the impact of deployment properties by using random byte payloads of size $256\,\mathrm{B}$ \cite{DBLP:conf/eurosys/DecouchantKRY22, DBLP:conf/ipps/DecouchantKRY24} and adding emulated network round trip latencies of $0\ms$ (same datacenter), $5\ms$ (same country), $35\ms$ (Europe), and $150\ms$ (World) to the physical datacenter round trip latency ($\sim 0.15\ms$).
We use the \nxb{} client model.
For each configuration and in a total of $\sim 2\mathrm{k}$ experiments, we identify the maximum request rate that saturates the system but does not cause overload: After experiment start, the latency emulation is started and the system is stressed with a fixed request rate.
We measure the achieved throughput in $\kops$ for $240\s$.
We follow a rather conservative rejection strategy: If request rate and throughput do not match or the system shows any burst or backlog patterns, the request rate is not accepted.
Please note that except for Chained-Damysus, a latency of $150\ms$, and $n\geq20$, we identify the maximum request rate with a granularity of $1\kops$.
Additionally, we measure the time between two decisions (intermediate decision time, IDT): 
In MinBFT, a replica is able to decide whenever it collects $f+1$ commits for a block (w/o faults one network round trip in expectation).
In Chained-Damysus, a replica is able to decide a block when it is followed by two valid and consecutive blocks (w/o faults one network round trip in expectation).
In \nxbft{}, a replica is able to decide when it finished a wave, i.e., it finished four consecutive broadcast rounds and the wave root selected by the common coin is part of the local DAG (w/o faults two network round trips in expectation).
\Cref{tab:network_peak} lists the maximum request rates in bold font and $\kops$ and the average IDT of 30 runs in normal font and milliseconds.
IDT confidence intervals are all below $9\%$ and left out.

MinBFT's throughput peaks for $n=3$ and no emulated latency with $69\kops$ and an IDT of $124\ms$.
MinBFT achieves the lowest throughput for $n=40$ and a network latency of $150\ms$. 
Chained-Damysus peaks for $n=3$ and no latency with $223\kops$ and an impressive IDT as small as $7\ms$.
For $n=40$ and a network latency of $150\ms$, the throughput drops to $0.2\kops$.
The IDT increases with network latency and $n$.
\nxbft{} shows peak performance for $n=10$ and no latency with $584\kops$ and $23\ms$ IDT. 
For network round trip latencies $\geq5\ms$, the throughput peaks for $n=3$.
For $n=40$ and $150\ms$ network round trip latency, the throughput drops to $20\kops$.
The IDT increases with network latency and $n$.

Although MinBFT requires at least one network round trip time between two decisions (two all-to-all broadcasts), MinBFT can stay below $150\ms$ IDT for $150\ms$ network round trip latency.
This is due to MinBFT's pipelining: the leader proposes a new block whenever it has sufficient requests. 
Thus, as long as the system is not overloaded, MinBFT can benefit from ABCperf's parallelized network stack.
This is in clear contrast to Chained-Damysus and \nxbft{} that both need quorums (which act as synchronization barriers) before being able to continue.
The throughput of MinBFT, however, cannot benefit: The network latency determines the minimum time a request has to wait for decision as the leader is required to collect a commit quorum.
To counter this, the leader could propose bigger blocks which would increase the time a replica requires to validate the block and compute its commit.
In both cases, too high request rates lead to maximally filled request queues at the side of the leader leading to dropped requests.

For network round trip latencies $\leq 5 \ms$, Chained-Damysus can take full advantage of the streamlined design, achieving IDTs clearly faster than MinBFT and \nxbft.
As observed before (cf.~Sec.~\ref{subsec:nxbpenalty} and \Cref{fig:exp:byzantine}), Chained-Damysus severely suffers from the \nxb{} client model when having big $n$. 
Block timeouts smaller than the selected $0.1\s$ may circumvent this behavior.
\nxbft{} shows the benefit of the \nxb{} client model in terms of load balancing, significantly outperforming throughput in all configurations.
While, for experiments with no payload, the load balancing allowed for higher throughput with increased $n$, both Chained-Damysus and \nxbft{} loose this ability.
\nxbft{}, however, keeps the ability for no added latency.
In fact, the load balancing increases the throughput but the increased cryptographic work, primarily driven by the payload size, masks the observed effect.

For Chained-Damysus and \nxbft, we observe that the IDT may decrease when increasing $n$, which indicates that the true maximum throughput may be higher.
First, our step size of $1\kops$ may be too big.
Second, larger networks may be more sensitive to scheduling and transmission variance when operating close to maximum performance.

\subsection{Performance Under Faults}

\begin{figure}
    \centering
    \includegraphics[width=\columnwidth]{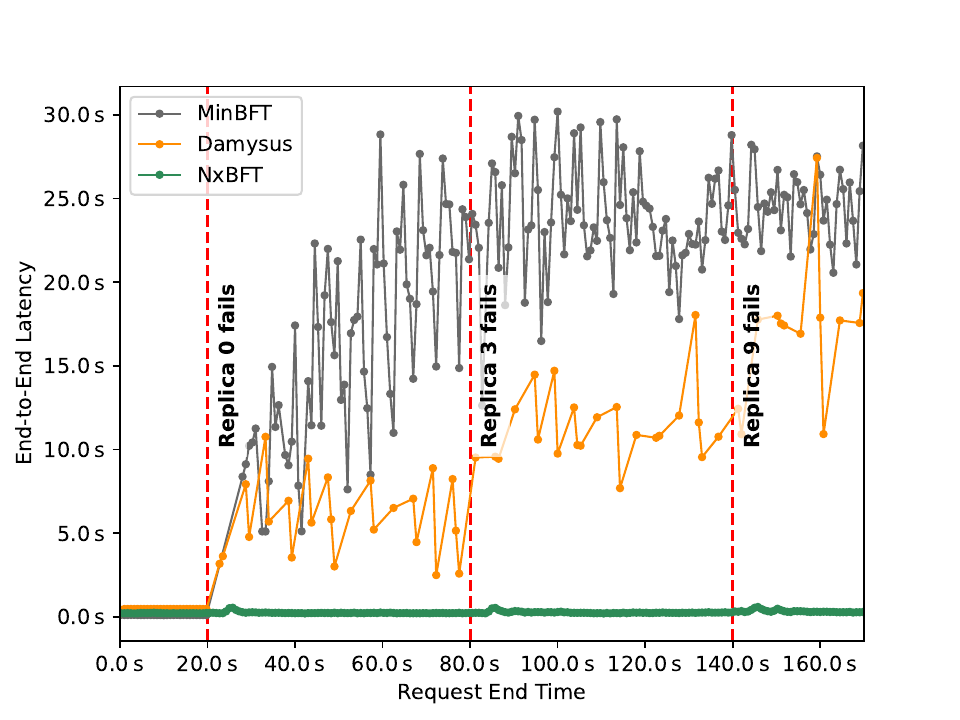}
    \caption{End-to-end-latency of MinBFT, Chained-Damysus, and \nxbft{} for  $n=10$ and a request rate of $50\kops$ under faults; average of ten runs. The plot shows the average end-to-end latency for requests that are successfully answered (if any) in a time buckets of size $0.75\,\mathrm{s}$. After $20\,\mathrm{s}$, $80\,\mathrm{s}$, and $140\,\mathrm{s}$, a replica fails. In case of MinBFT, the crash of replica 0 causes a view change.}
    \label{fig:exp:fault}
\end{figure}

We investigate the algorithms' performance under faults for $n=10$, a request rate of $50\kops$ $0\mathrm{B}$ payloads using the \nxb{} client model, and no emulated latency.
An experiment lasts $400\,\mathrm{s}$; $230\,\mathrm{s}$ at the start are left for warm-up.
After $20\,\mathrm{s}$, replica 0 crashes, after $80\,\mathrm{s}$, replica 3 crashes, and after $140\,\mathrm{s}$, replica 9 crashes.
We selected those replicas to crash as this pattern allows Chained-Damysus to make the most progress.
To prevent unlimited increase of Damysus' timeouts, we do not crash four replicas (circumvention requires orthogonal consensus ~\cite{Malkhi2022Latest}).
\Cref{fig:exp:fault} shows the average end-to-end latency of the experiment for ten repetitions.

In case of MinBFT, replica 0 is the current leader and MinBFT is forced to perform a view-change.
The throughput of MinBFT stalls until the inactivity of the leader is detected~(i.e., $5\,\mathrm{s}+3\,\mathrm{s}$) and the view-change was successful.
The new leader is not capable to cut the backlog down, yielding increased latencies for the remainder of the experiment. 
The crash of non-leader replicas (3 and 9) has no effect.
For lower request rates, MinBFT would be capable of recovering fast response times within the experiment time window.

Without a recovery procedure, Chained-Damysus cannot recover the response times and each crashed replica increases the end-to-end latency clearly.
Since Chained-Damysus requires a replica to be in charge for two views \cite[App. B]{Decouchant2022DamysusExtended}, the streamlined version of Damysus worsens this pattern.

As all client requests are uniformly distributed across all replicas, \nxbft{}'s maximum achievable throughput degrades at most by $\frac{c}{n}$ for $c$ crashed replicas.
For the request rate of $50\kops$, our experiments show that the $7$ remaining replicas can handle the additional load:
As soon as all clients identified a failed replica, the end-to-end latency recovers.

\subsection{\texorpdfstring{\nxbft{} Recovery Protocol}{NxBFT Recovery Protocol}}

\begin{table}[t!]
\renewcommand{\arraystretch}{1.1}
\caption{\nxbft{} Recovery Times Depending on Handled Request Count\\($256\,\mathrm{B}$ Payload, $0.15\ms$ Network Round Trip Latency)}
\label{tab:recovery}
\centering
\begin{tabular}{l|c|c|c}
       & $125\mathrm{k}$ & $250\mathrm{k}$ & $500\mathrm{k}$ \\ 
\hline 
$n=3$  & $0.55\pm 0.02\s$       & $2.30\pm0.07\s$        & $8.57\pm0.95\s$ \\
$n=10$ & $0.29\pm 0.00\s$       & $0.89\pm0.05\s$        & $3.71\pm0.36\s$ \\
$n=20$ & $0.34\pm0.01\s$        & $0.66\pm0.03\s$        & $2.35\pm0.15\s$ \\
$n=40$ & $0.43\pm0.01\s$        & $0.79\pm0.02\s$        & $2.45\pm0.07\s$ 
\end{tabular}
\end{table}

We investigate the recovery time of crashed \nxbft{} replicas for $n\in\{3,10,20,40\}$ and request rates of $25\kops$, $50\kops$, and $100\kops$ using the \nxb{} client model. 
After $5\s$, a replica is crashed and recovered.
In this time, the system handled a total of $125\mathrm{k}$, $250\mathrm{k}$, and $500\mathrm{k}$ client requests.
We measure the time it takes for the crashed replica to be able to propose valid messages.
\Cref{tab:recovery} lists the average recovery times with the $95\%$ confidence interval for 30 repetitions. 
Our experiment shows that the recovery time is primarily influenced by the number of vertices a replica sent before it crashed.
The number of messages is influenced by the time until the replica crashes, the request rate and the number of replicas $n$.
The request rate determines how many messages a crashed replica produced, $n$ controls load balancing in the \nxb{} client model.
Regular checkpointing would limit the graph history and, thus, speedup recovery significantly.

\section{Discussion and Future Work}
\label{sec:discussion}

\nxbft's performance achievements are built upon the assumptions regarding possible attacker behavior: The \nxb{} operating model allows reducing load between clients and replicas. Load balancing in general scales  with the number of replicas, however, the quadratic overhead of the broadcast layer eventually becomes the dominating factor limiting the maximum number of replicas.
The missing resilience against attacks from operators on clients can be mitigated by the application signing responses.
In case of such an attack, a client can prove its correct behavior and the client state can be corrected.
Instead, one could also deploy the application to a TEE as well or use off-loading technologies like CART~\cite{Hess2024Consensus}.

Operators and other third parties could attack the TEE.
Such an attack would not leak any additional application data in comparison to classic BFT SMR and requires, e.g., secure multi-party computation for mitigation.
On the consensus layer, however, knowing coin values beforehand increases the chance for successful censorship \cite{danezis2022narwhal}, and on the broadcast layer equivocation could bring the system to a halt.
We consider the probability of attacks on consensus and broadcast to be negligible in the intended deployment scenarios.

Setup and recovery require some form of synchrony to achieve liveness.
We argue that, while being unconventional for secure distributed systems research, the proposed model and assumptions are well-suited for practical deployments.
Federations and consortia built upon \nxbft{} will operate a highly resilient common service.
Maintenance and recovery windows can be agreed upon beforehand and peers can recover without interrupting the service. 
While then formally resulting in a partially synchronous system, \nxbft{} operates asynchronously once set up or recovered.

As future work, gar\-bage collection is required: With thousands of operations per second, the required memory for storing the graph grows quickly.
A garbage collection that actually deletes information, e.g., based on decided waves as in \cite{danezis2022narwhal}, however, is in conflict with asynchrony \emph{and} the backfilling strategy described above: 
If a replica learns rather ``late'' (in relation to other replicas) that it misses some information, other replicas may have reached consensus and deleted this information already. 
The wave counter is a local variable and gives no sufficient information on the synchronization progress with other replicas.
Scheduled maintenance time slots, as discussed above, ``trivially'' facilitate garbage collection: 
Our recovery protocol can be extended to derive a full system checkpoint, but, still, depending on all replicas to cooperate.
From there on, implementing a voting-based reconfiguration scheme becomes easy as well.

The wave length of four rounds in \nxbft{} is required to prove the atomic broadcast's agreement property \cite[Proposition 3]{DBLP:conf/podc/KeidarKNS21} based on the \emph{get core} property \cite[Lemma 4]{DBLP:conf/wdag/LeinweberH23}. 
The get core property is required to derive an expectation value for a correct replica being able to commit a wave, i.e., a liveness guarantee.
Related work \cite{danezis2022narwhal, DBLP:journals/corr/abs-2501-01062} proposed to shorten the wave length to three rounds.
However, with a wave length of only three rounds and a fault tolerance of $n > 2f$, the get core property does not hold anymore (see Appendix~\ref{sec:get_core_rounds}).

As our results confirm, partially synchronous protocols typically trade good latency for smaller throughput and overhead in the case of faults.
Within the last two years, parallel work optimized this trade-off by adopting ideas of leaderless and DAG-based protocols in partial synchrony \cite{DBLP:journals/corr/MalkhiSY23, DBLP:journals/iacr/ShresthaKN24, DBLP:conf/sosp/GiridharanSAAC24, DBLP:journals/corr/abs-2405-20488}. 
Furthermore, researchers achieved impressive performance results for asynchronous protocols in non-Byzantine environments \cite{DBLP:conf/eurosys/WangLDW0ZZ024}.
Based on their promising results, we consider the investigation of the integration of TEEs and fault model relaxations to be a promising line of future research.


\section{Conclusion}
\label{sec:conclusion}

We presented and evaluated \nxbft, a full-fledged TEE-based State Machine Replication approach in the ``Not eXactly Byzantine'' operating model. 
\nxbft{} bases its safety and liveness properties of normal case operation solely on the assumption of asynchrony and is, therefore, highly resilient.
Setup and recovery, however, require some form of synchrony.
Using graph-based agreement and load balancing, \nxbft{} demonstrates competitive throughput performance while maintaining a reasonable latency trade-off.
In our opinion, further optimization and increased deployability of DAG-based approaches that use TEEs represent interesting future work.  

\section*{Acknowledgments}
This work was supported by funding from the topic Engineering Secure Systems of the Helmholtz Association (HGF). 
We would like to thank Tilo Spannagel for his expertise during the implementation work and Oliver Stengele for his feedback on setup and recovery protocols.

\bibliographystyle{IEEEtran}
\bibliography{references.bib}

\appendix

\subsection{Correctness Arguments}
\label{sec:correctness}

\subsubsection{Backfilling-Based Reliable Broadcast}

In the following, we will argue that the backfilling mechanism as described in Sec.~\ref{subsubsec:broadcast} transforms the best effort broadcast to a reliable broadcast.
Since we use the signature service to implement a FIFO broadcast with non-equivocation, we get `no duplication', `integrity', and `consistency' as reliable broadcast properties \cite[Module 3.12]{DBLP:books/daglib/CachinGR2011} for free. 
`Validity' and `totality' properties require that if a correct replica broadcasts or delivers a message, every other correct replica will deliver this message; both are based on the backfilling mechanism:
Correct replicas will only use vertices as edges that they already added to their DAG. 
This behavior implies that those edges are valid vertices with a completely known ancestry.
If a replica now receives a vertex for which it does not know an edge, this is due to two reasons: (1) the edge is from a correct replica and the corresponding vertex was not yet delivered by the asynchronous channel or (2) the edge is from a faulty replica that committed a send omission fault.
In case (1), our replica will receive the edge eventually and validity and totality will be preserved.
In case (2), if at least one correct replica received the vertex and was able to add it to the DAG, no matter if itself used the vertex as an edge or not, the replica will be able to answer any vertex request, thus, fulfilling totality.
If no correct replica received the vertex, no correct replica will deliver the vertex, thus, not breaking totality.

\subsubsection{Common Coin}

The XOR construction is a variant of the straightforward $t=n$ secret sharing; the resulting seed is kept confidential by secure communication and the TEE.
The enclave will not reveal a toss unless at least one correct replica requests a toss.
A correct replica will always toss a coin when it completes the last round of a wave and proposed its own vertex for it, enabling every correct replica to learn the coin value as soon as possible.
Note that it is not necessary to use the TEE-based signatures, as we do, to achieve the desired properties, but it makes implementation easier.

\subsubsection{Recovery Protocol}

Obviously, the proposed recovery protocol is only live if all replicas are eventually reactive:
Byzantine faulty replicas can stop the recovery procedure by not sending a RecoveryProposal at any time (or an invalid one).
In the following, we will argue that the proposed recovery protocol is Byzantine safe in an asynchronous environment.
To prevent that the recovery protocol enables equivocation, the enclaves of all correct replicas must replace the recovering replica $p_c$'s enclave key with the same new enclave public key (i.e., agreement on the $p_c$'s new attestation certificate).
To this end, a correct replica collects $\quorum > r$ valid RecoveryCommits that it passes to its enclave.
The enclave verifies the signatures and only accepts if all $\quorum$ RecoveryCommits are correctly signed.
As a correct await $\quorum>f$ RecoveryCommits, a correct replica will receive at least one RecoveryCommit created by a correct replica.
Moreover, as a correct replica waits for $n$ RecoveryProposals with consistent attestation certificates before sending a RecoveryCommit, the enclave logic can deduce agreement on the supplied attestation certificate and rule out equivocation of the new enclave public key of the recovering replica.
Since RecoveryProposal and RecoveryCommit result in a reliable broadcast of proposals, all correct replicas apply the exact same vertices of a recovering replica $p_c$ when they identify the longest valid vertex chain.
The common coin is recovered by importing the enclave state, including the seed established during setup, and fast-forwarding the PRNG state to be directly ``before'' the next toss the replica would have made when not crashing.
Thus, the coin implementation will only reveal new coins whenever the replica can prove the needed progress of the system (with or without the replica's contribution) to its enclave.
As the common coin uses the public enclave keys, the agreement on the attestation certificates ensures that a Byzantine faulty replica cannot learn coin values in advance.

\subsection{Impossibility of Get Core with a Reduced Wave Length}
\label{sec:get_core_rounds}

Graph-based atomic broadcast protocols of the DAG-Rider family rely for their liveness on ``connectivity guarantees'': After a certain amount of rounds, there is the guarantee that \emph{all} valid vertices of a round $r' > r$ have a path to a subset of the vertices of round $r$.
A vertex is valid if it is correctly signed, it links to at least $2f+1$ (for BFT) or $f+1$ (for hybrid fault models) vertices of the previous round, and its ancestry is known, valid, and part of the local DAG.
The original DAG-Rider approach \cite{DBLP:conf/podc/KeidarKNS21} requires that when selecting four consecutive and completed rounds $r_1, r_2, r_3, r_4$ \emph{all} vertices of $V_4$ of round $r_4$ have a path to the \emph{same} subset $V_{\mathrm{core}}$ of round $r_1$.
Furthermore, it is required that this common subset, or common core, is at least of size $|V_{\mathrm{core}}| \geq 2f+1$.
This property can be proven by using the \emph{get core} scheme developed by Attiya and Welch \cite[Sec. 14.3.1]{attiya2004distributed} (sometimes also known as ``Gather'').
TEE-Rider showed that the property still holds when using a wave length of four, i.e., four consecutive rounds, a size of $|V_{\mathrm{core}}| \geq f+1$, and a fault tolerance of $n>2f$ if it can be assumed that each possible malicious action of an attacker can be reduced to an omission fault \cite[Lemma 4]{DBLP:conf/wdag/LeinweberH23}.

Intuition would suggest that by using TEEs the wave length could be reduced in a similar way as TEEs save a round of communication for PBFT-like protocols \cite{veronese2013efficient}.
In \Cref{fig:get_core_counter}, we constructed a counter example for a wave length of $3$ and $n=5$ breaking \emph{get core}.
The \emph{get core} property requires that each valid vertex links to at least $3$ vertices of the previous round. 
Every vertex is valid and its ancestry is part of the graph. 
However, the blue vertices (i.e., the vertices of round 3) only share a subset of round 1 with size $2$ (the green vertices labeled 0 and 1).
If \emph{get core} would hold for three rounds, there should be a shared subset of size $3$.
This is the case if a fourth round of valid vertices connecting to the blue vertices is added.
If a liveness proof does not rely on \emph{get core}, it seems possible to shorten the wave length.
However, to the best of our knowledge, no such liveness proof for an asynchronous, graph-based, and Byzantine fault tolerant atomic broadcast protocol with a fault tolerance of $n>2f$ is known yet.

\begin{figure}
    \centering
\begin{tikzpicture}[->, node distance=1cm]
\node[circle, draw, fill=red!50] (4) {4};
\node[circle, draw, fill=red!50] (3) [below of=4] {3};
\node[circle, draw, fill=red!50] (2) [below of=3] {2};
\node[circle, draw, fill=green!50] (1) [below of=2] {1};
\node[circle, draw, fill=green!50] (0) [below of=1] {0};
\node[circle, draw, fill=gray!50] (9) [right of=4, node distance=3cm] {9};
\node[circle, draw, fill=gray!50] (8) [right of=3, node distance=3cm] {8};
\node[circle, draw, fill=gray!50] (7) [right of=2, node distance=3cm] {7};
\node[circle, draw, fill=gray!50] (6) [right of=1, node distance=3cm] {6};
\node[circle, draw, fill=gray!50] (5) [right of=0, node distance=3cm] {5};
\node[circle, draw, fill=blue!50] (14) [right of=9, node distance=3cm] {14};
\node[circle, draw, fill=blue!50] (13) [right of=8, node distance=3cm] {13};
\node[circle, draw, fill=blue!50] (12) [right of=7, node distance=3cm] {12};
\node[circle, draw, fill=blue!50] (11) [right of=6, node distance=3cm] {11};
\node[circle, draw, fill=blue!50] (10) [right of=5, node distance=3cm] {10};

\path (9) edge (4);
\path (9) edge (3);
\path (9) edge (0);
\path (8) edge (3);
\path (8) edge (1);
\path (8) edge (0);
\path (7) edge (2);
\path (7) edge (1);
\path (7) edge (0);
\path (6) edge (4);
\path (6) edge (1);
\path (6) edge (0);
\path (5) edge (2);
\path (5) edge (1);
\path (5) edge (0);
\path (14) edge (9);
\path (14) edge (8);
\path (14) edge (6);
\path (13) edge (8);
\path (13) edge (7);
\path (13) edge (5);
\path (12) edge (7);
\path (12) edge (6);
\path (12) edge (5);
\path (11) edge (7);
\path (11) edge (6);
\path (11) edge (5);
\path (10) edge (7);
\path (10) edge (6);
\path (10) edge (5);
\end{tikzpicture}
    \caption{Counter example for the \emph{get core} property with $n=5$ and a wave length of three rounds. All blue vertices (vertices of round 3) should have a path to a shared subset of round 1 (red and green vertices) of size 3. The maximum shared subset, however, is of size 2 (green vertices).}
    \label{fig:get_core_counter}
\end{figure}

\end{document}